\documentclass{www2006-submission}
\usepackage{graphicx}
\def\more-auths{%
\end{tabular}
\begin{tabular}{c}}
\def\sharedaffiliation{%
\end{tabular}
\begin{tabular}{c}}
\begin{document}
\title{The egalitarian effect of search engines}
\numberofauthors{2}

\author{
\alignauthor Santo Fortunato$^{1,2}$\\
       \email{santo@indiana.edu}
\alignauthor Alessandro Flammini$^{1}$\\
       \email{aflammin@indiana.edu}
\vskip0.5pc  
\more-auths
\alignauthor Filippo Menczer$^{1}$\\
       \email{fil@indiana.edu}
\alignauthor Alessandro Vespignani$^{1}$\\
       \email{alexv@indiana.edu}
\vskip1pc  
\sharedaffiliation
\affaddr{$^{1}$ School of Informatics}  \\
\affaddr{Indiana University}   \\
\affaddr{Bloomington, IN 47406, USA}
\sharedaffiliation
\affaddr{$^{2}$ Fakult\"at f\"ur Physik}  \\
\affaddr{Universit\"at Bielefeld}   \\
\affaddr{D-33501 Bielefeld, Germany}
}
\date{}

\maketitle

\begin{abstract}
Search engines have become key media for our scientific, economic, and
social activities by enabling people to access information on the Web
in spite of its size and complexity.  On the down side, search engines
bias the traffic of users according to their page-ranking strategies,
and some have argued that they create a vicious cycle that amplifies
the dominance of established and already popular sites.  We show that,
contrary to these prior claims and our own intuition, the use of search
engines actually has an egalitarian effect.  We reconcile theoretical
arguments with empirical evidence showing that the combination of
retrieval by search engines and search behavior by users mitigates the
attraction of popular pages, directing more traffic toward less popular
sites, even in comparison to what would be expected from users randomly
surfing the Web.
\end{abstract}

\category{H.3.3}{Information Storage and Retrieval}{Information Search and Retrieval}
\category{H.3.4}{In\-for\-ma\-tion Storage and Retrieval}{Systems and Software}[Information networks]
\category{H.3.5}{In\-for\-ma\-tion Storage and Retrieval}{Online Information Services}[Commercial, Web-based services]
\category{H.5.4}{In\-for\-ma\-tion Interfaces and Presentation}{Hypertext/Hypermedia}[Navigation, user issues]
\category{K.4.m}{Com\-puters and Society}{Miscellaneous}

\terms{Measurement}

\keywords{Search engines, bias, popularity, traffic, PageRank, in-degree.}

\section{Introduction}

The crucial role of the Web as a communication medium and its
unsupervised, self-organized development have triggered the intense
interest of the scientific community.  The topology of the Web as a
complex, scale-free network is now well
characterized~\cite{Albert99,Kleinberg99short,Broder00,Adamic00,Kleinberg01}.
Several growth and navigation models have been proposed to explain the
Web's emergent topological characteristics and their effect on users'
surfing
behavior~\cite{Barabasi99,Kumar00,Kleinberg00,Pennock02,Menczer02navigation,Menczer03attach,Barrat04traffic}.
As the size and complexity of the Web have increased, users have become
reliant on search engines~\cite{Lawrence98,Lawrence99}, so that the
paradigm of search is replacing that of navigation as the main
interface between people and the Web~\cite{websidestory,Cho05WebDB}.
This leads to questions about the role of search engines in shaping the
use and evolution of the Web.

\begin{figure*}
\begin{center}
\centerline{\includegraphics[width=\textwidth]{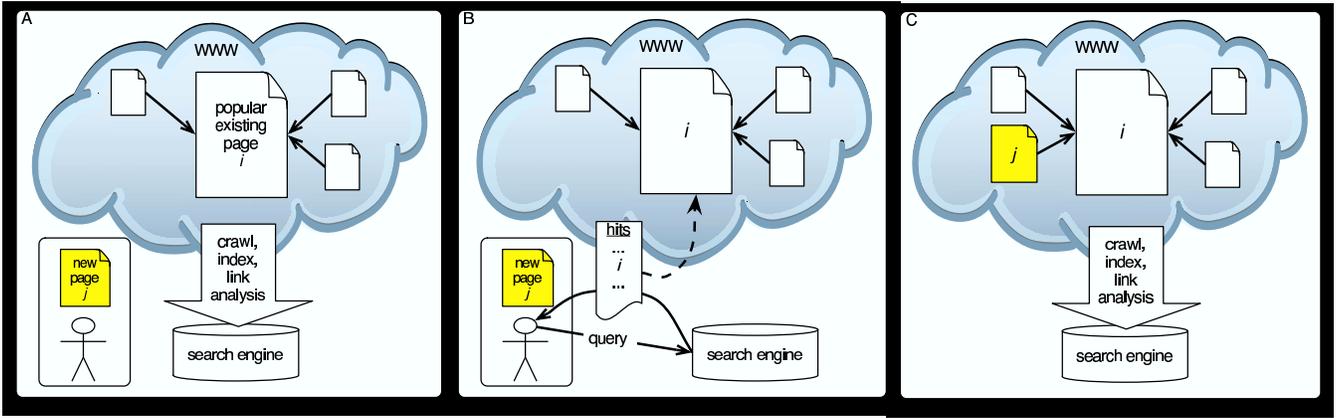}}
\caption{Illustration of search engine bias.
\textbf{A}. Page $i$ is ``popular'' in that it has many incoming links and high PageRank. 
A user creates a new page $j$.
\textbf{B}. The user consults a search engine to find pages related to $j$.  Since $i$ is ranked highly by the search engine, it has a high probability of being returned to the user.
\textbf{C}. The user, having discovered $i$, links to it from $j$.  Thus $i$ becomes even more popular from the search engine's perspective.}
\label{fig0}
\end{center}
\end{figure*}

One common belief is that the use of search engines biases 
traffic toward popular sites.  This is at the origin of the
vicious cycle illustrated in Fig.~\ref{fig0}.  Pages highly ranked by
search engines are more likely to be discovered and consequently 
linked to by other pages.  This in turn would further increase 
the popularity and raise the average rank of
those pages.  As popular pages become more and more popular, new
pages are unlikely to be discovered~\cite{Cho04impact}.  Such a cycle
would accelerate the rich-get-richer dynamics already observed in the
Web's network structure and explained by preferential attachment and
link copy models~\cite{Barabasi99,Kleinberg99short,Kumar00}. This presumed 
phenomenon, also known as \emph{search engine bias, entrenchment effect,} 
or \emph{googlearchy}, has been widely discussed in computer, social and 
political science~\cite{Introna00SearchPolitics,Mowshowitz02Bias,Baeza-Yates02,Hindman03googlearchy,Cho04impact,PandeyRoyOlstonChoChak05VLDB} 
and methods to counteract it are being 
proposed~\cite{Cho05SIGMOD,PandeyRoyOlstonChoChak05VLDB}. 

In this paper we use both empirical and theoretical arguments to 
show that the bias of search engines is of the opposite nature, 
namely directing more traffic toward less popular pages compared 
to the case in which no search occurs and all traffic is generated 
by surfing hyperlinks. Our contributions are organized as follows:

\begin{itemize}

\item We develop a simple modeling framework in which one can 
quantify the amount of traffic that Web sites receive 
in the extreme cases in which users browse the Web by surfing 
random hyperlinks and in which users only visit pages returned 
by search engines in response to queries. The framework, introduced 
in Section~\ref{sect2}, allows to make and compare predictions 
about how navigation and search steer traffic and thus bias the 
popularity of Web sites.

\newpage
\item In Section~\ref{sect3} we provide a first empirical study 
of the traffic toward Web pages as a function of their in-degree. 
This particular relationship is the one that can directly validate 
the models in Section~\ref{sect2}. As it turns out, both the surfing 
and searching models are surprisingly wrong; the bias in favor 
of popular pages seems to be mitigated, rather than enhanced, by 
the combination of search engines and users' search behavior. 
This result contradicts prior assumptions about search engine bias.

\item The unexpected empirical observation on traffic is explained in 
Section~\ref{sect4}, where we take into consideration a previously 
neglected factor about search results, namely the distribution and 
composition of 
hit set size. This distribution, determined empirically from actual 
user queries, allows one to reconcile the searching model with the
empirical data of Section~\ref{sect3}. Using theoretical 
arguments and numerical simulations we show that the search 
model, revised to take queries into account, accurately predicts 
traffic trends confirming the egalitarian bias of search engines.

\end{itemize}

\section{Modeling the vicious cycle}
\label{sect2}

For a quantitative definition of popularity we turn to the probability
that a generic user clicks on a link leading to a specific
page~\cite{Cho05SIGMOD}.  We will also refer to this quantity as the
traffic to the same page.

\subsection{Surfing model of traffic}
 
In the absence of search engines, people would browse Web pages
primarily by following hyperlinks.  It is natural to assume that the
amount of such surfing-generated traffic directed toward a given page
is proportional to the number of links $k$ pointing to it.  The more
the pages pointing to that page, the larger the probability that a
randomly surfing user will discover it.  Successful search engines,
Google being the premier example~\cite{Brin98}, have modeled this
effect in their ranking functions to gauge page importance.  The
PageRank value $p(i)$ of page $i$ is defined as the probability that a
random walker moving on the Web graph will visit $i$ next, thereby
estimating the page's discovery probability according to the global
structure of the Web.  Experimental observations and theoretical
results show that, with good approximation, $p \sim k$ (see
Appendix~\ref{A1}).  Therefore, in the surfing model where users only
visit pages by following links, the traffic through a page is given by
$t \sim p \sim k$.

\subsection{Searching model of traffic}
 
When navigation is mediated by search engines, to estimate the traffic
directed toward a page, one must consider how search engines retrieve
and rank results, as well as how people use these results.  Following
the seminal paper by Cho and Roy~\cite{Cho04impact}, 
this means that we need to find two
relationships: (i) how the PageRank translates into the rank of a
result page, and (ii) how the rank of a hit translates into the
probability that the user clicks on the corresponding link thus
visiting the page. 

The first step is to determine the scaling relationship between
PageRank (and equivalently in-degree as discussed above) and rank.
Search engines employ many factors to rank pages.  Such factors are
typically query-dependent: whether the query terms appear in the title
or body of a page, for example.  They also use a global
(query-independent) importance measure, such as PageRank, to judge the
value of search hits.  If we average across many user queries, we
expect PageRank to determine the average rank $r$ of each page within
search results: the page with the largest $p$ has average rank $r
\simeq 1$ and so on, in decreasing order of $p$.

\begin{figure}[tb!]
\centerline{\includegraphics[width=\columnwidth]{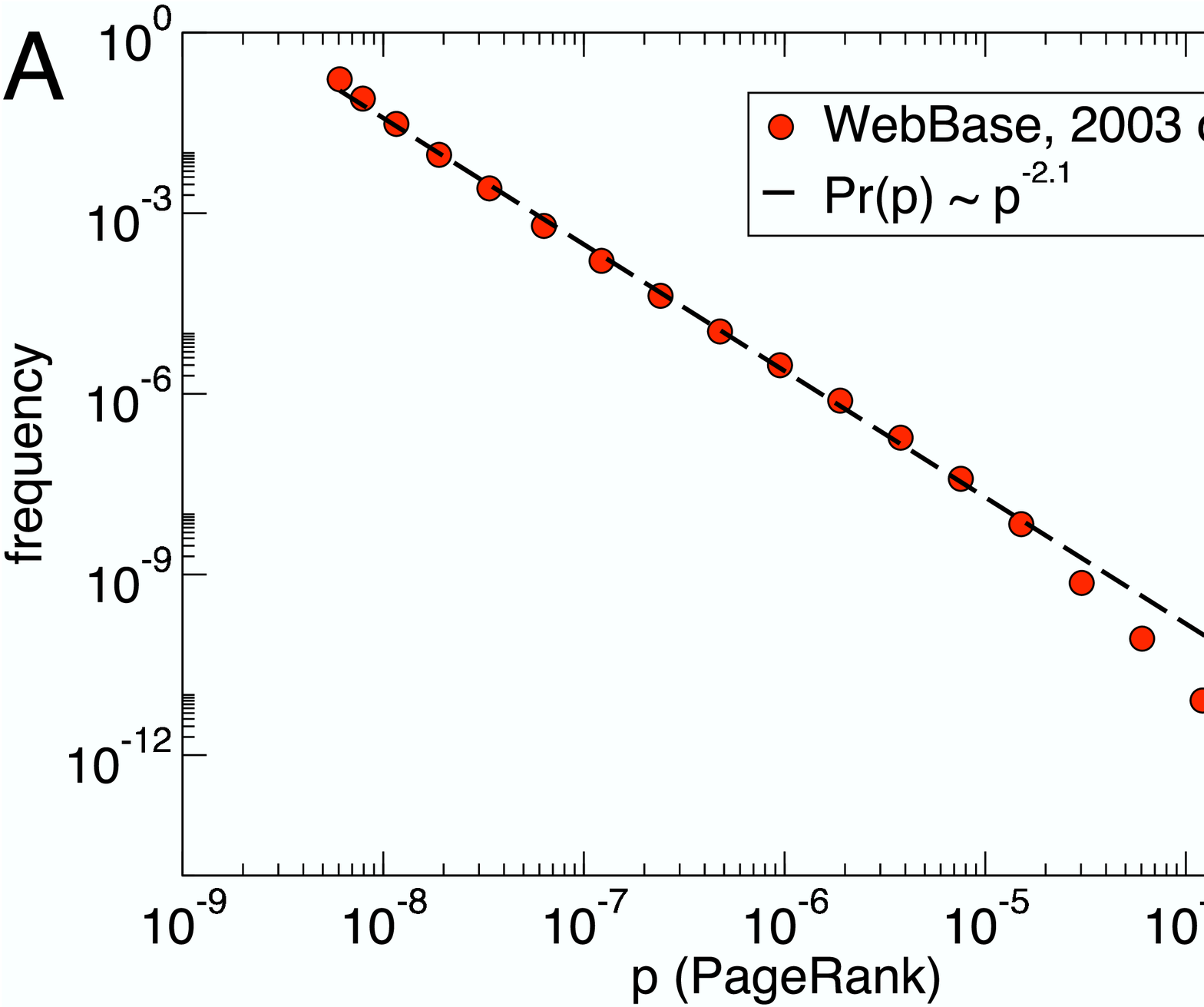}}
\centerline{\includegraphics[width=\columnwidth]{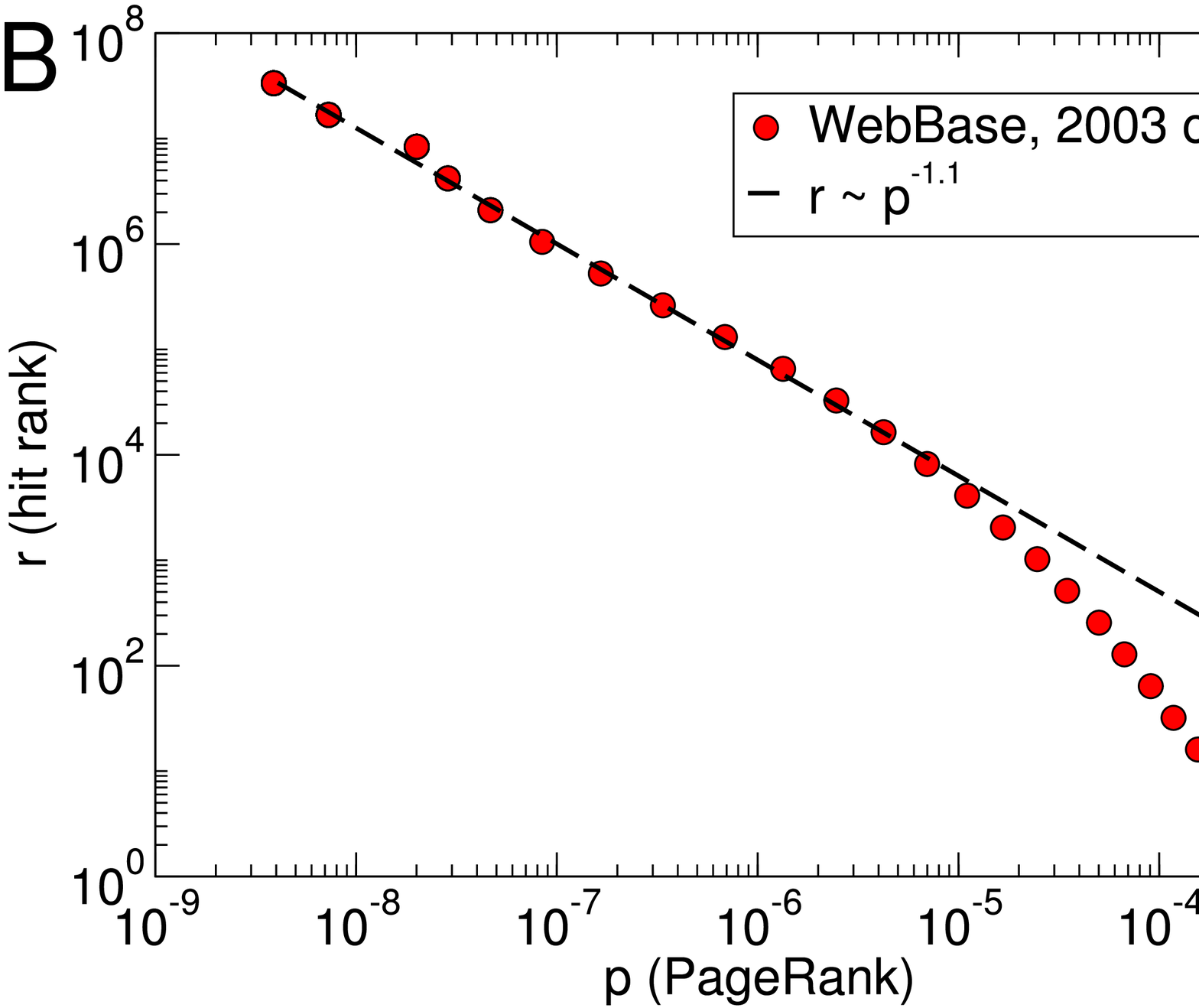}}
\caption{\label{figsuppl03}
\textbf{A}: Distribution of PageRank $p$: the log-log plot shows 
a power law   $\Pr(p) \sim p^{-2.1}$.
\textbf{B}:~Empirical relation between rank and PageRank: the log-log 
plot shows a power law   $r \sim p^{-1.1}$.
Both plots are based on data from a WebBase 2003 
crawl~\protect\cite{WebBase}.
}
\end{figure}

Statistically, $r$ and $p$ have a non-linear relationship.  
There is an exact mathematical relationship between the value of a
variable $p$ and the rank of that value, assuming that a set of
measures is described by a normalized histogram (or distribution)
$Pr(p)$.  The rank $r$ is essentially the number of measures greater
than $p$, i.e., $r = N \int_p^{p_{max}} Pr(x) dx$, where $p_{max}$ is
the largest measure gathered and $N$ the number of measures.  
Empirically we find that the distribution of PageRank is a power 
law $p^{-\mu}$ with exponent $\mu \approx 2.1$ (Fig.~\ref{figsuppl03}A). 
In general, when the variable $p$ is distributed according 
to a power law with exponent $-\mu$ and
neglecting large $N$ corrections one obtains:
\begin{equation}
r(p) \sim p^{-\beta}
\label{eq10}
\end{equation}
where $\beta=\mu - 1 \approx 1.1$. 
Cho and Roy~\cite{Cho04impact} derived the relation between $p$ and $r$
differently, by fitting the empirical curve of rank vs.  PageRank
obtained from a large WebBase crawl.  Their fit returns a somewhat
different value for the exponent $\beta$ of $3/2$.  To check this
discrepancy we used Cho and Roy's method and fitted the empirical curve
of rank vs.  PageRank from our WebBase sample, confirming our estimate
of $\beta$ over three orders of magnitude (Fig.~\ref{figsuppl03}B).

\begin{figure}
\begin{center}
\includegraphics[width=\columnwidth]{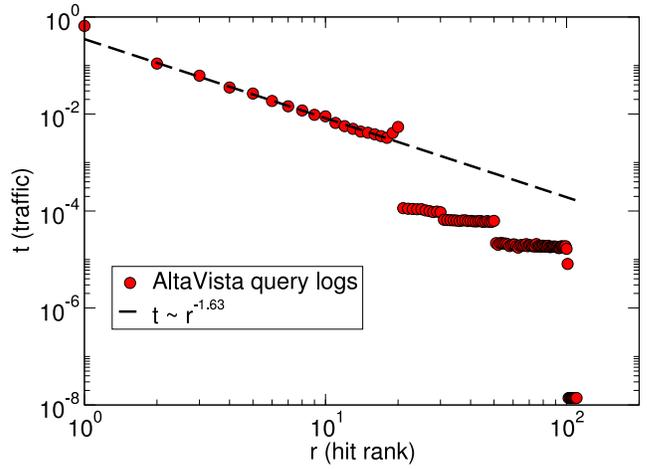}
\caption{
Scaling relationship between click probability $t$ and hit rank $r$: the log-log plot shows a power law $t \sim r^{-1.63}$ 
(data from a sample of 7 million queries submitted to AltaVista between September 28 and October 3, 2001). 
}
\label{fig1}
\end{center}
\end{figure}

The second step, still following ref.~\cite{Cho04impact}, is to
approximate the traffic to a given page by the probability that when
the page is returned by a search engine, the user will click on its
link.  We expect the traffic $t$ to a page to be a decreasing function
of its rank $r$.  Lempel and Moran~\cite{LempelMoran03} reported a
non-linear relation $t \sim r^{-\alpha}$, confirmed by our analysis
using query logs from AltaVista as shown in Fig.~\ref{fig1}.

\newpage Note that the rank plotted on the x-axis of Fig.~\ref{fig1} does not
refer exactly to the absolute position of a hit $i$ in the list of
hits, but rather to the rank of the result page where the link to $i$
appears.  Search engines display query results in pages containing a
fixed number of hits (usually 10).  Assuming that each result page
contains 10 items, as in the Altavista queries we examined, all hits
from the first to the tenth will appear in the first result page and
the corresponding click probabilities will be cumulated, giving the
leftmost point in the plot.  The same is done for the hits from the
$11^{th}$ to the $20^{th}$, from the $21^{st}$ to the $30^{th}$, and so
on.  In lack of better information we consider result pages instead of
single hits, implicitly assuming that within each result page the
probability to click on a link is independent of its position.  This
assumption is reasonable, although there can still be a gradient
between the top and the bottom hits, as people usually read the list
starting from the top.

The sudden drop near the $21^{st}$ result page in Fig.~\ref{fig1} is
due to the way AltaVista operated during the summer 2001, when they
decided to limit the list of results to 200 pages per query (displayed
in 20 result pages).  We therefore limited the analysis to the first 20
data points, which can be fitted quite well by a simple power law
relation between the probability $t$ that a user clicks on a hit and
the rank $r_{p}$ of the result page where this hit is displayed:
\begin{equation}
t \sim {r_p}^{-\alpha}
\label{eq5}
\end{equation}
with exponent $\alpha = 1.63 \pm 0.05$.  The fit exponent obtained by
Cho and Roy was $3/2$, which is close to our estimate.

In our calculations we took into account the grouping of the hits in
result pages, consistently with the empirical result of
Fig.~\ref{fig1}.  However we noticed that if one replaces in
Eq.~\ref{eq5} the rank $r_p$ of the result page with the absolute rank
$r$ of the individual hits, the final results do not change
appreciably.  Therefore to simplify the discussion we shall assume from
now on that
\begin{equation}
t \sim {r}^{-\alpha}.
\label{eq6}
\end{equation}
The rapid decrease of $t$ with the rank $r$ of the hit clearly
indicates that users focus with larger probability on the top results.

We are now ready to express the traffic as a function of page in-degree
$k$ using the general scaling relation $t \sim k^{\gamma}$.  In the
pure surfing model, $\gamma = 1$; in the searching model, we take
advantage of the relations between $t$ and $r$, between $r$ and $p$,
and between $p$ and $k$ to obtain 
\begin{equation}
t \sim r^{-\alpha} \sim (p^{-\beta})^{-\alpha} = p^{\alpha \beta} \sim k^{\alpha \beta}  
\label{eq10b}
\end{equation}
and therefore $\gamma = \alpha \beta$, ranging between $\gamma \approx
1.8$ (according to our measures $\alpha \approx 1.63$, $\beta \approx
1.1$) and 2.25 (according to estimates by others~\cite{LempelMoran03,
Cho04impact}).  

In all cases, the searching model leads to a value $\gamma>1$.  This
superlinear behavior implies that the common use of search engines will
bias traffic toward already popular sites.  This is at the origin of
the vicious cycle illustrated in Fig.~\ref{fig0}.  Pages highly ranked
by search engines are more likely to be discovered (as compared to pure
surfing) and consequently linked to by other pages.  This in turn would
further increase their PageRank and raise the average rank of those
pages.  Popular pages become more and more popular, while new pages are
unlikely to be discovered~\cite{Cho04impact}.  Such a cycle would
accelerate the rich-get-richer dynamics already observed in the Web's
network structure~\cite{Barabasi99,Kleinberg99short,Kumar00}.  This
presumed phenomenon has been dubbed \emph{search engine bias} or
\emph{entrenchment effect} and has been recently brought to the 
attention of the technical Web community~\cite{Baeza-Yates02,Cho04impact,PandeyRoyOlstonChoChak05VLDB},
and methods to counteract it have been
proposed~\cite{Cho05SIGMOD,PandeyRoyOlstonChoChak05VLDB}.  There are
also notable social and political implications to such a
\emph{googlearchy}~\cite{Introna00SearchPolitics,Mowshowitz02Bias,Hindman03googlearchy}.

\section{Empirical data}
\label{sect3}

\begin{figure}[t!]
\centerline{\includegraphics[width=\columnwidth]{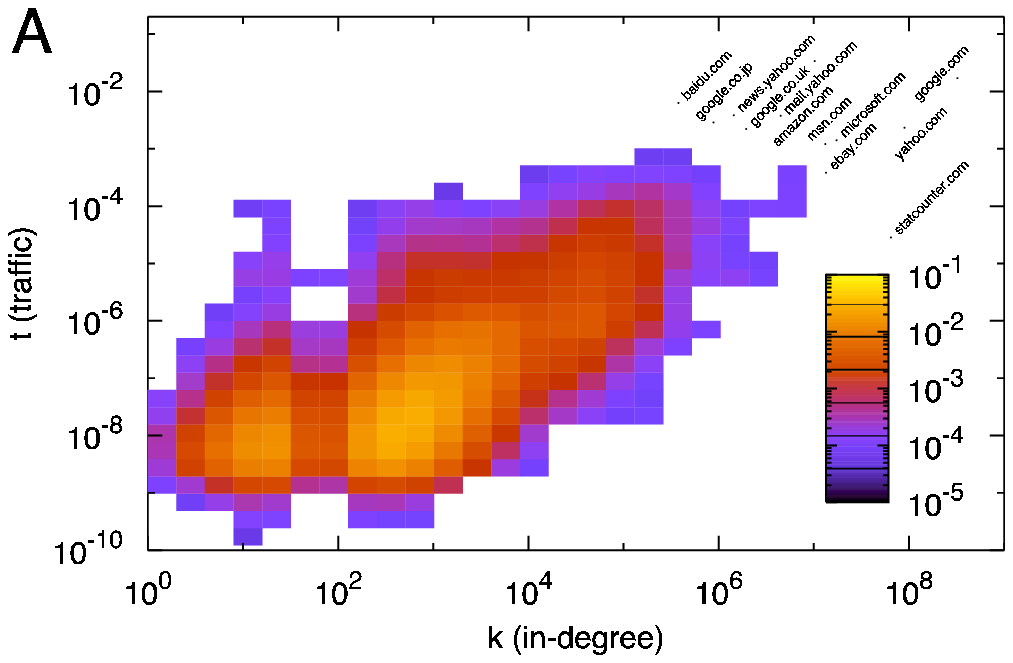}}
\centerline{\includegraphics[width=\columnwidth]{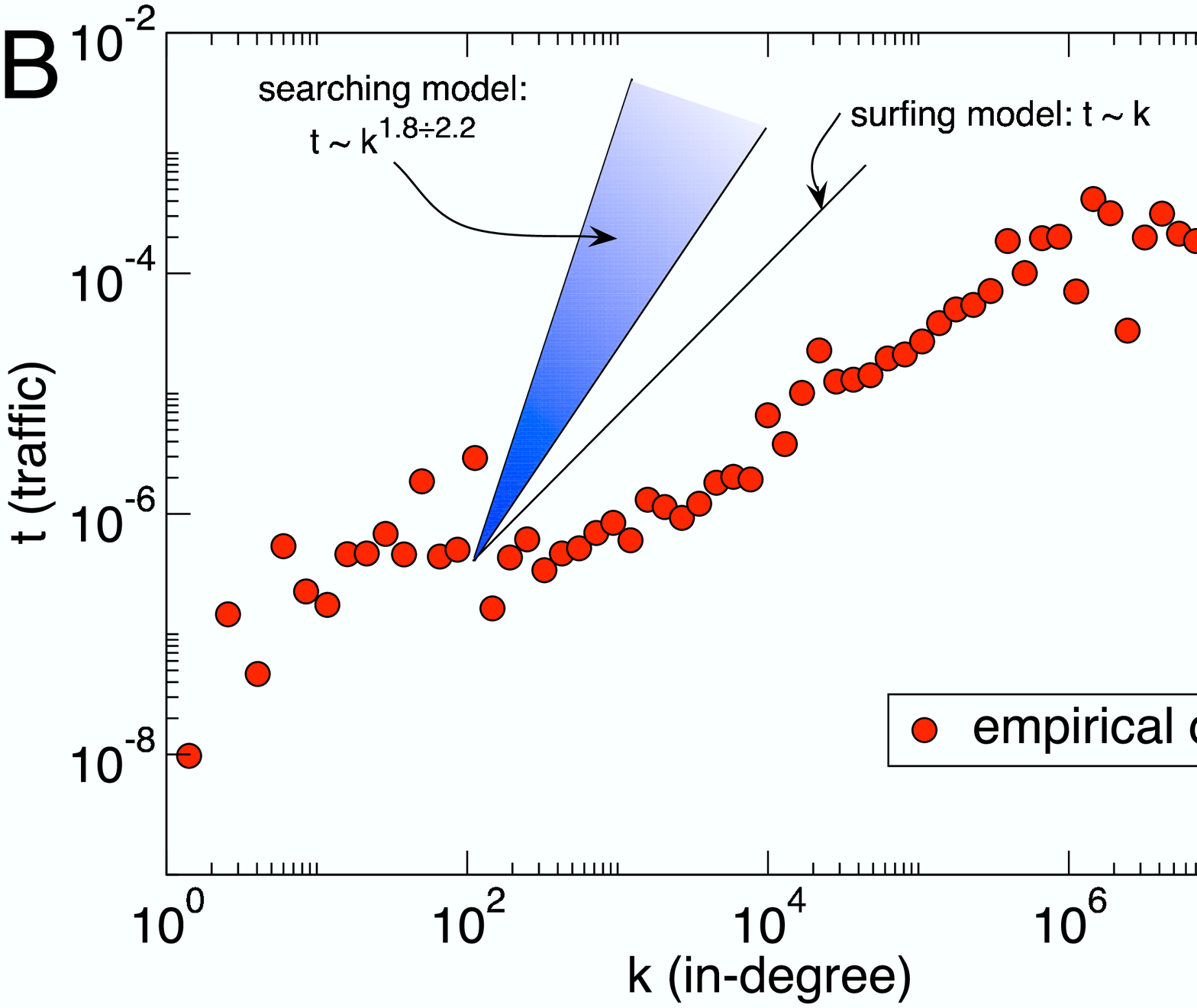}}
\caption{\textbf{A}. Density plot of traffic versus in-degree for a sample of 28,164 Web sites.  Colors represent the fraction of sites in each log-size bin, on a logarithmic color scale.  A few sites with highest in-degree and/or traffic are highlighted.  The source of in-degree data is Yahoo~\protect\cite{YahooAPI}; using Google~\protect\cite{GoogleAPI} yields the same trend.  Traffic is measured as the fraction of all page views in a three-month period, according to Alexa data~\protect\cite{alexa}.   
\textbf{B}. Relationship between average traffic and in-degree obtained with logarithmic binning of in-degree.  The power-law predictions of the surfing and searching models discussed in the text are also shown.}
\label{fig2}
\end{figure}

To determine whether such a vicious cycle really exists, let us
consider the empirical data.  Given a Web page, its in-degree is the
number of links pointing to it, which can be easily estimated using a
search engine such as Google or Yahoo~\cite{GoogleAPI,YahooAPI}.
Traffic is the fraction of all user clicks in some period of time that
lead to the page; this quantity, also known as \emph{view
popularity}~\cite{Cho05SIGMOD}, can be estimated using the Alexa
Traffic Rankings service, which monitors the sites viewed by users of
its toolbar~\cite{alexa}.  We used the Yahoo and Alexa services to
estimate in-degree and traffic for a total of 28,164 Web pages.  Of
these, 26,124 were randomly selected using Yahoo's random page service.
The remaining 2,040 pages were selected among the sites that Alexa
reports as the ones with highest traffic.  The resulting density plot
is shown in Fig.~\ref{fig2}A.

To ensure the robustness of our analysis, we collected our data twice
at a distance of two months.  While there were differences in the
numbers (for example Yahoo increased the size of its index
significantly in the meanwhile), there were no differences in the
scaling relations.  We also collected in-degree data using
Google~\cite{GoogleAPI}, again yielding different numbers but the same
trend.  The in-degree measures exclude links from the same site.  For
example, to find the in-degree for \texttt{\small
http://informatics.indiana.edu/}, we would submit the query
``\texttt{link:http://informatics.indiana.edu/} \linebreak
\texttt{-site:informatics.indiana.edu}''.  Note that the in-degree data
provided by search engines is only an estimate of the true number.
First, a search engine can only know of links from pages that it has
crawled and indexed.  Second, for performance reasons, the algorithms
counting inlinks use various unpublished approximations based on
sampling.

Traffic is measured as page views per million in a three-month period.
Alexa collects and aggregates historical traffic data from millions of
Alexa Toolbar users.  Page views measure the number of pages viewed by
these users.  Multiple page views of the same page made by the same
user on the same day are counted only once.  Our measure of 
traffic $t$ corresponds to Alexa's count, divided by $10^{6}$ to 
express the fraction of all the page views by toolbar users go to a
particular site.  Since traffic data is only available for Web sites
rather than single pages, we correlate the traffic of a site with the
in-degree of its main page.  For example, suppose that we want the
traffic for \texttt{\small http://informatics.indiana.edu/}.  Alexa
reports the 3-month average traffic of the domain \texttt{\small
indiana.edu} as 9.1 page views per million.  Further, Alexa reports
that 2\% of the page views in this domain goes to the \linebreak 
\texttt{\small
informatics.indiana.edu} subdomain.  Thus we reach the estimate of
0.182 page views per million.

To derive a scaling relation, we average traffic along logarithmic bins
for in-degree, as shown in Fig.~\ref{fig2}B. Surprisingly, both the
searching and surfing models fail to match the observed scaling, which
is not modeled well by a power law.  Contrary to our expectation, the
scaling relation is sublinear, suggesting that search engines actually
have an \textit{egalitarian} effect, directing more traffic than
expected to less popular sites --- those having lower PageRank and
fewer links to them.  Search engines thus have the effect of
counteracting the skewed distribution of links in the Web, directing
some traffic toward sites that users would never visit otherwise.  This
result is at odds with the previous theoretical discussion; in order to
understand the empirical data, we need to include a neglected but basic
feature of the Web: the semantic match between queries and page content.

\section{Queries and hit set size}
\label{sect4}

\begin{figure}
\centerline{\includegraphics[width=\columnwidth,height=2.3in]{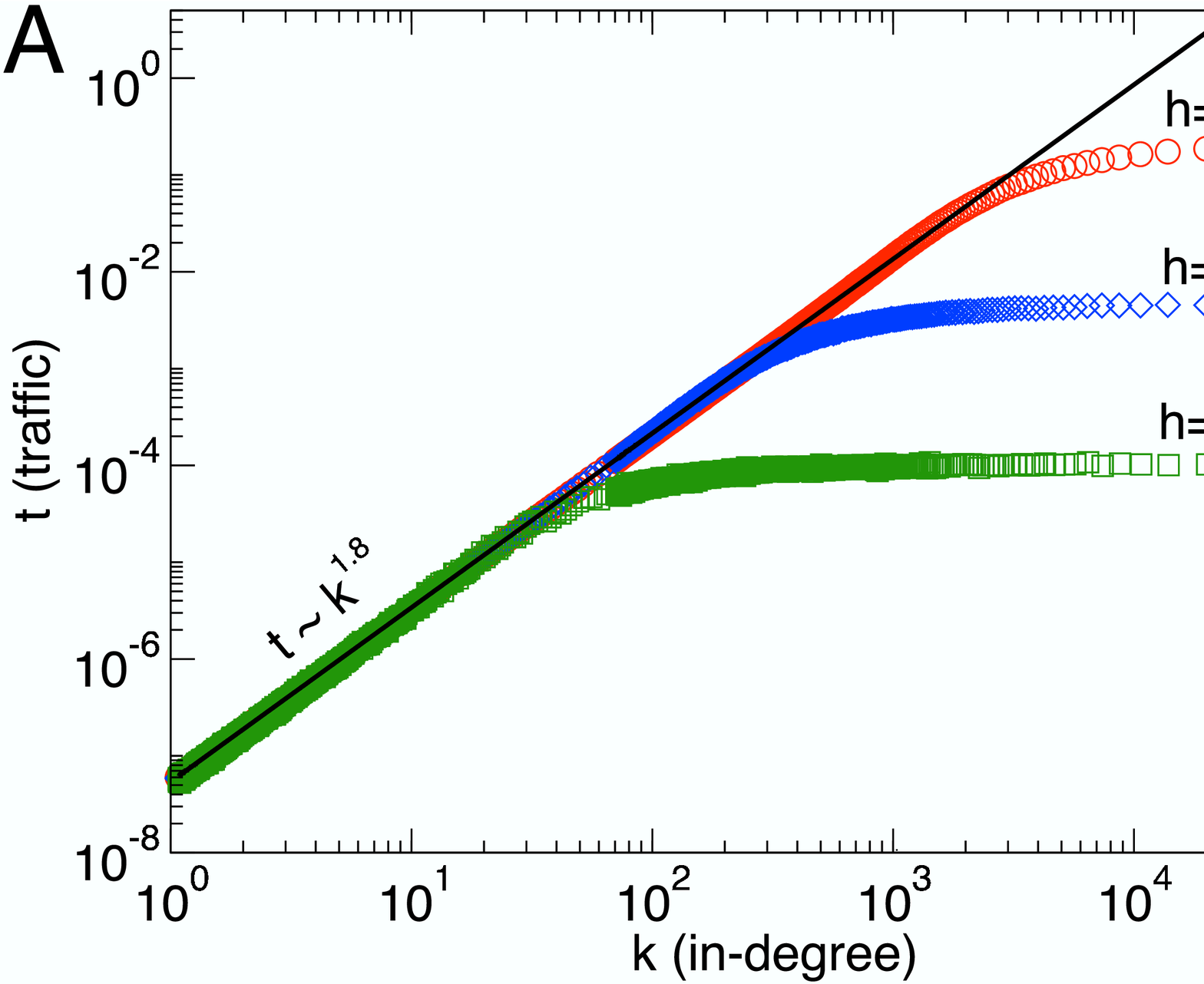}} 
\centerline{\includegraphics[width=\columnwidth,height=2.3in]{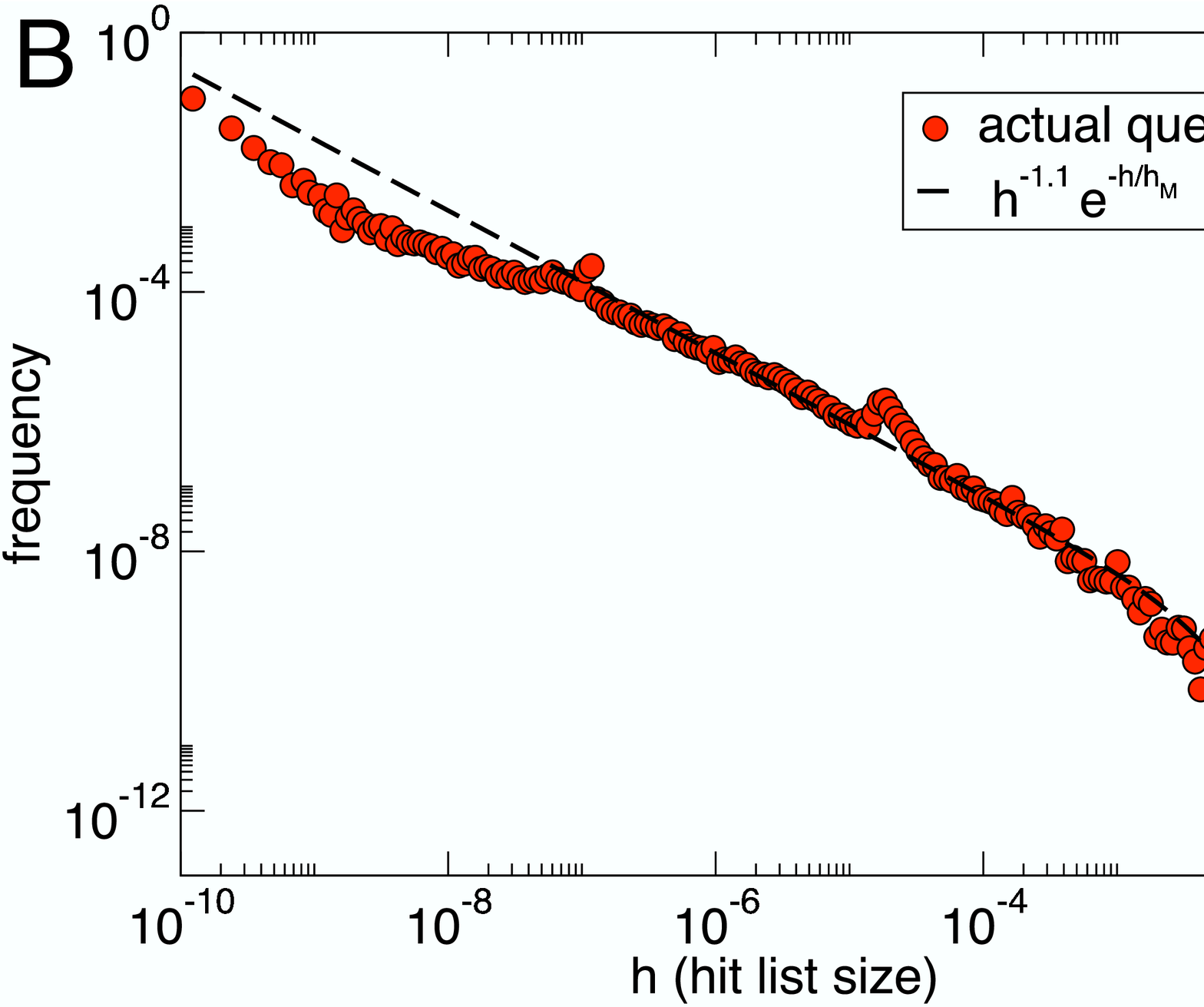}}
\centerline{\includegraphics[width=\columnwidth]{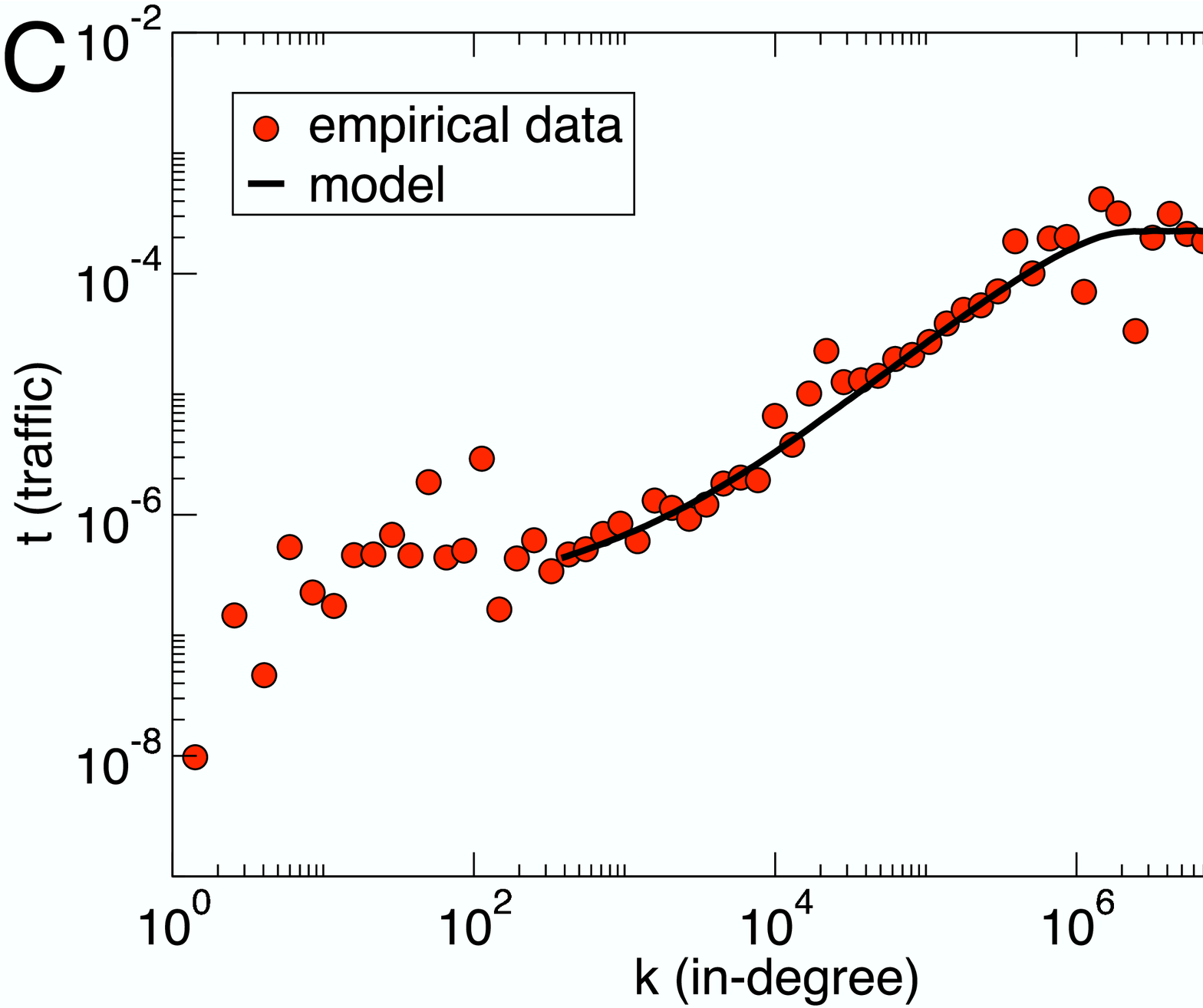}}
\caption{\textbf{A}. Scaling relationship between traffic and in-degree when each page has a fixed probability $h$ of being returned in response to a query. 
The curves (not normalized for visualization purposes) are obtained by simulating the process $t[r(k),h]$ (see Appendix~\ref{A2}). 
\textbf{B}. Distribution of relative hit set size $h$ for 200,000 actual user queries from AltaVista logs.  The hit set size data were 
obtained from Google~\protect\cite{GoogleAPI}.  Frequencies are normalized by logarithmic bin size.  The log-log plot shows a power 
law with an exponential cutoff. 
\textbf{C}. Scaling between traffic and in-degree obtained by simulating 4.5 million 
queries with a realistic distribution of hit set size on a one-million node network. 
Empirical data from Fig.~\ref{fig2}B.
} 
\label{fig3}
\end{figure}

In the previous theoretical estimate of traffic as driven by search
engines, we considered the global rank of a page, computed across
\textit{all} pages indexed by the search engine.  However, any given
query typically returns only a small number of pages compared to the
total number indexed by the search engine.  The size of the ``hit''
set and the nature of the query introduce a significant bias in the
sampling process.  If only a small fraction of pages are returned in
response to a query, their rank within the set is not representative
of their global rank as induced, say, by PageRank.  

Let us assume that all query result lists derive from a
Bernoulli process such that the number of hits relevant 
to each query is on average $hN$ where $h$ is the 
relative hit set size. In Appendix~\ref{A2} 
we show that this assumption leads to an alteration in the 
relationship between traffic and in-degree. 
To illustrate this effect, Fig.~\ref{fig3}A shows how the click
probability changes with $h$.  The result $t \sim k^{\gamma}$ 
(or $t \sim r^{-\alpha}$, cf.  Fig.~\ref{fig1}) only holds in the
limit case $h \rightarrow 1$.  Since the size of the hit sets is not
fixed, but depends on user queries, we measured the distribution of hit
set sizes for actual AltaVista queries as shown in Fig.~\ref{fig3}B,
yielding $\Pr(h) \sim h^{-\delta}$, with $\delta \approx 1.1$ over 
seven orders of magnitude. The exponential cutoff in the distribution 
of $h$ is due to the maximum size $h_{M}$ of actual hit lists 
corresponding to non-noise terms, and thus can be disregarded for 
our analysis.

The traffic behavior is therefore a convolution of the different curves
reported in Fig.~\ref{fig3}A, weighted by $\Pr(h)$. 
The final relation between traffic and
degree can thus be obtained by numerical techniques (see
Appendix~\ref{A2}) and, strikingly, the resulting behavior reproduces
the empirical data over four orders of magnitude, including the
peculiar saturation observed for high-traffic sites (Fig.~\ref{fig3}C).
Most importantly, the theoretical behavior predicts a traffic increase
for pages with increasing in-degree that is noticeably slower than the
predictions of both the surfing and searching models.  In other words,
the combination of search engines, the semantic attributes of queries,
and users' own behavior mitigates the rich-get-richer dynamics of the
Web, providing low-degree pages with increased visibility.

Of course, actual Web traffic is a combination of both surfing and
searching behaviors.  Users rely on search engines heavily, but also
navigate from page to page through static links as they explore the
neighborhoods of pages returned in response to search
queries~\cite{Cho05WebDB}.  It would be easy to model a mix of our
revised searching model (taking into account the more realistic
distribution of hit set sizes) with the random surfing behavior.  The
resulting mixture model would yield a prediction somewhere between the
linear scaling $t \sim k$ of the surfing model (cf.  Fig.~\ref{fig2}B)
and the sublinear scaling of our searching model (cf.
Fig.~\ref{fig3}C).  The final curve would be sublinear and still in
agreement with the empirical traffic data.

\section{Discussion and outlook}
\label{sect5}

Our heavy reliance on search engines as a means of coping with the
Web's size and growth does affect how we discover, link to, and visit
pages.  However, in spite of the rich-get-richer dynamics implicitly
contained in the use of link analysis to rank search hits, the net
effect of search engines on traffic appears to produce an egalitarian
effect, smearing out the traffic attraction of high-degree pages.  Our
empirical data clearly shows a sublinear scaling relation between
referral traffic from search engines and page in-degree.  This seems to
be in agreement with the observation that search engines lead users to
visiting about 20\% more pages than surfing alone~\cite{Cho05WebDB}.
Such an effect may be understood within a theoretical model of
information retrieval that considers the users' clicking behavior 
and the heavy-tailed distribution observed for the number of query 
hits.

This result has relevant conceptual and practical consequences.  It
suggests that, contrary to intuition and prior hypotheses, the use of
search engines contributes to a more level playing field, in which new
Web sites have a greater chance of being discovered and thus of
acquiring links and popularity --- as long as they are about specific
topics that match the interests of users as expressed through their
search queries.  

Such a finding is particularly relevant for the design of realistic
models for Web growth.  The connection between the popularity of a page
and its acquisition of new links has led to the well-known
rich-get-richer growth paradigm that explains many of the observed
topological features of the Web.  The present findings, however, show
that several non-linear mechanisms involving search engine algorithms
and user behavior regulate the popularity of pages.  This calls for a
new theoretical framework that considers more of the various behavioral
and semantic issues that shape the evolution of the Web.  How such a
framework may yield coherent models that still agree with the Web's
observed topological properties is a difficult and important
theoretical challenge.

Finally, the present results provide a first quantitative estimate of,
and prediction for, the popularity and traffic generated by Web pages.
This estimate promises to become an important tool to be exploited in
the optimization of marketing campaigns, the generation of traffic
forecasts, and the design of future search engines.

\section{Acknowledgments}

We thank the members of the Networks and Agents Network at
IUB, especially Mark Meiss, for helpful feedback on early 
versions of the manuscript. 
We are grateful to Alexa, Yahoo and Google for extensive use of
their Web services, to the Stanford WebBase project for their crawl
data, and to AltaVista for use of their query logs.  
This work is funded in part by a Volkswagen Foundation grant to SF, 
by NSF awards 0348940 and 0513650 to FM and AV respectively, and by 
the Indiana University School of Informatics.

\appendix
\section{Relationship between \\ in-degree and PageRank}
\label{A1}

Let us inspect the scaling relationship 
between in-degree $k$ and PageRank $p$. 
In our calculations of PageRank we used a damping factor 
$0.85$, as in the original version of the 
algorithm~\cite{Brin98} and in many successive studies.
Our numerical analysis of the PageRank for the Web graph 
was performed on two samples
produced by crawls made in 2001 and 2003 by the WebBase 
collaboration at Stanford \cite{WebBase}.
The graphs are quite large: the former crawl has 
80,571,247 pages and 752,527,660 links; the latter  
has 49,296,313 pages and 1,185,396,953 links. 

\begin{figure}
    \centerline{\includegraphics[width=\columnwidth]{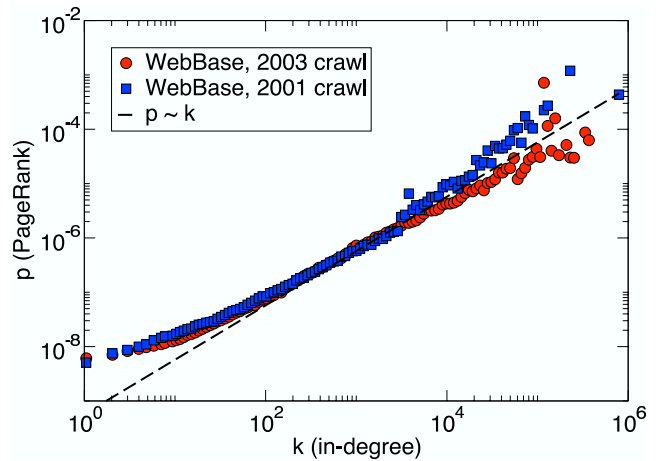}} 
    \caption{\label{figsuppl02}
    PageRank as a function of in-degree for two samples 
    of the Web taken in 2001 and 2003 \protect\cite{WebBase}.}
\end{figure}

In Fig.~\ref{figsuppl02}, in order to reduce fluctuations, 
we averaged the PageRank values over 
logarithmic bins of the degree. 
The data points mostly fall on a power law curve for both samples, 
with $p$ increasing with $k$. 
The correlation coefficients of the two sets of data, before binning, 
are $0.54$ and $0.48$ for the 2001 and 2003 crawl, respectively, 
as found for the Web domain of the University of Notre Dame~\cite{naka}, 
but in disagreement with the results of an analysis on the domain of Brown University 
and the WT10g Web snapshot~\cite{brown}.
The estimated exponents of the power law fits for the two curves 
are $1.1 \pm 0.1$ (2001) and $0.9 \pm 0.1$ (2003). 
As shown in Fig.~\ref{figsuppl02}, the two estimates are 
compatible with a simple linear relation between PageRank and in-degree.
A linear scaling relation between $p$ and $k$ is also consistent with 
the observation that both have the same distribution. 
As it turns out, $p$ and $k$ are both distributed 
according to a power law with estimated exponent $-2.1 \pm 0.1$, 
in agreement with other estimates~\cite{brown,millozzi,Broder00}.  
We assume, therefore, that PageRank and in-degree are, on average, 
proportional for large values.

\section{Simulation of search-driven \\ Web traffic}
\label{A2}

When a user submits a query to a search engine, the latter will select
all pages deemed relevant from its index and display the corresponding
links ranked according to a combination of query-dependent factors,
such as the similarity between the terms in the query and those in the
page title, and query-independent prestige factors such as PageRank.
Here we focus on PageRank as the main global ranking factor, assuming
that query-dependent factors are averaged out across queries.  The
number of hit results depends on the query and it is in general much
smaller than the total number of pages indexed by the search engine.

Let us start from the relation between click probability and rank in
Eq.~\ref{eq6}.  If all $N$ pages in the index were listed in each
query, as implicitly assumed in ref.~\cite{Cho04impact}, the
probability for the page with the smallest PageRank to be clicked would
be $N^{\alpha}$ ($\alpha \approx 1.63$ in our study) times smaller than
the probability to click on the page with the largest PageRank.
If instead both pages ranked first and $N^{th}$ appear among the $n$
hits of a realistic query (with $n \ll N$), they would still occupy the
first and the last positions of the hit list, but the ratio of their
click probabilities would be much smaller than before, i.e.
$n^{\alpha}$.  This leads to a redistribution of the clicking
probability in favor of the less ``popular'' pages, which are then
visited much more often than one would expect at first glance.
To quantify this effect, we must first distinguish between the
\emph{global} rank induced by PageRank across all Web pages and the
\emph{query-dependent} rank among the hits returned by the search
engine in response to a particular query.  Let us rank all $N$ pages in
decreasing order of PageRank, such that the global rank is $R=1$ for
the page with the largest PageRank, followed by $R=2$ and so on.

Let us assume for the moment that all query result lists derive from a
Bernoulli process with success probability $h$ (i.e., the number of
hits relevant to each query is on average $hN$).  The assumption that
each page can appear in the hit list with the same probability $h$ is
in general not true, as there are pages that are more likely to be
relevant than others, depending on their size, intrinsic appeal, and so
on.  If one introduces a fitness parameter to modulate the probability
for a page to be relevant with respect to a generic query, the results
would be identical as long as the fitness is not correlated with the
PageRank of the page.  In what follows we then stick to the simple
assumption of equiprobability.

Let us calculate the probability $\Pr(R,r,N,n,h)$ that the page with
global rank $R$ has rank $r$ within a list of $n$ hits.  This is the
probability $p_{r-1}^{R-1}$ to select $r-1$ pages from the set $\{1
\dots R-1\}$:
\begin{eqnarray}
p_{r-1}^{R-1} &=& h^{r-1}(1-h)^{R-1-(r-1)}\left(\begin{array}{c}R-1 \\
r-1 \\
\end{array} \right) \nonumber \\
&=& h^{r-1}(1-h)^{R-r}\left(\begin{array}{c}R-1 \\
r-1 \\
\end{array}\right)
\label{eq12}
\end{eqnarray}
times the probability $p_{n-r}^{N-R}$ to select
$n-r$ pages from the set $\{R+1 \dots N\}$, 
times the probability $h$ to select page $R$. So we obtain:
\begin{eqnarray}
\lefteqn{\Pr(R,r,N,n,h) = p_{r-1}^{R-1} p_{n-r}^{N-R} h} \nonumber \\
&=& h^n(1-h)^{N-n}
\left(\begin{array}{c}R-1 \\
r-1 \end{array} \right)
\left(\begin{array}{c}N-R \\
n-r \end{array} \right).
\label{eq14}
\end{eqnarray}
If page $R$ has rank $r$ in a list of $n$ hits, the probability of being clicked will be
\begin{equation}
t(R,r,N,n,h) = \frac{r^{-\alpha}}{\sum_{m=1}^n m^{-\alpha}} \Pr(R,r,N,n,h)
\label{eq15}
\end{equation}
where the denominator ensures the proper normalization of the click
probability within the hit list.  What remains to be done is to sum
over the possible ranks $r$ of page $R$ in the hit list ($r \in 1 \dots
n$) and over all possible hit set sizes ($n \in 1 \dots N$).  The final
result for the probability $t(R,N,h)$ of the $R$-th page to be clicked
is:
\begin{eqnarray}
\lefteqn{t(R,N,h) =} \nonumber \\ 
& \sum_{n=1}^{N}\sum_{r=1}^{n} & \frac{r^{-\alpha}}{\sum_{m=1}^{n} m^{-\alpha}} h^n(1-h)^{N-n} \cdot \nonumber \\
& & \cdot \left(\begin{array}{c}R-1 \\
r-1 \end{array} \right)
\left(\begin{array}{c}N-R \\
n-r \end{array} \right).
\label{eq20}
\end{eqnarray}

From Eq.~\ref{eq20} we can see that if $h=1$, which corresponds to a
list with all $N$ pages, one recovers Eq.~\ref{eq6}, as expected.  For
$h<1$, however, it is not possible to derive a close expression for
$t(R,N,h)$, so one has to calculate the binomials and perform the sums
numerically.  This can be easily done, but the time required to perform
the calculation increases dramatically with $N$, so that it is not
realistic to push the computation beyond $N=10^4$.  For this reason,
instead of carrying on an exact calculation, we performed Monte Carlo
simulations of the process leading to Eq.~\ref{eq20}.

In each simulation we produce a large number of hit lists, where every
list is formed by picking each page of the sample with probability $h$.
At the beginning of the simulation we initialize all entries of the
array $t(R,N,h)=0$.  Once a hit list is completed, we add to the
entries of $t(R,N,h)$, corresponding to the pages of the hit list, the
click probability as given by Eq.~\ref{eq6} (with the proper
normalization).  With this Monte Carlo method we simulated systems with
up to $N=10^6$ items.  To eliminate fluctuations we averaged the click
probability in logarithmic bins, as already done for the experimental
data.

\begin{figure}
    \centerline{\includegraphics[width=\columnwidth]{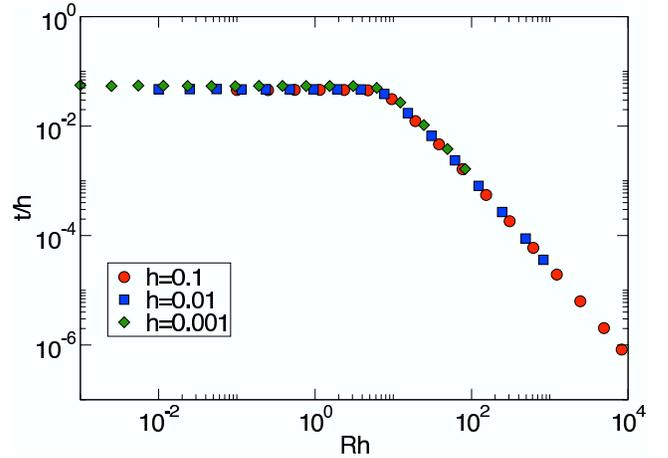}}
    \caption{\label{figsuppl7}
    Scaling of $t(R,N,h)/h$ with the variable $Rh$. The three curves refer to 
a sample of $N=10^5$ pages.}
\end{figure}

We found that the function $t(R,N,h)$ obeys a simple scaling law:
\begin{equation} 
t(R,N,h) = h\,F(Rh)A(N)
\label{eq22a}
\end{equation}
where $F(Rh)$ has the following form:
\begin{equation} 
F(Rh) \sim \left\{ 
  \begin{array}{ll}
    const & \mbox{if $h \leq Rh \leq 1$}\\
    (Rh)^{-\alpha} & \mbox{if $Rh \geq 1$}.
  \end{array} \right.
\label{eq22}
\end{equation}
An immediate implication of Eq.~\ref{eq22a} is that if one plots \linebreak $t(R,N,h)/h$
as a function of $Rh$, for $N$ fixed, one obtains the same curve $F(Rh)A(N)$, 
independently of the value of $h$ (Fig.~\ref{figsuppl7}).

The decreasing part of the curve $t(R,N,h)$, for $Rh>1$ i.e. $R>1/h$,
is the same as in the case when $h=1$ (Eq.~\ref{eq6}).  This means that
the finite size of the hit list affects only the top-ranked $1/h$
pages.  The effect is thus strongest when the fraction $h$ is small,
i.e., for specific queries that return few hits.  The striking feature
of Eq.~\ref{eq22} is the plateau for all pages between the first and
the $1/h$-th.  This implies that the difference in the values of
PageRank among the top $1/h$ pages does not
produce a difference in the probability of clicking on those pages. 
For $h=1/N$, which would correspond to lists containing on average a
single hit, each of the $N$ pages would have the same probability of
being clicked, regardless of their PageRank.  This is not surprising,
as we assumed that all pages have the same probability to appear in a
hit list.

So far we assumed that the number of query results is drawn from a
binomial distribution with a mean of $hN$ hits.  On the other hand, we
know that real queries generate a broad range of possible hit set
sizes, going from lists with only a single result to lists containing
tens of millions of results.  If the size of the hit list is
distributed according to some function $S(h,N)$, one would need to
convolute $t(R,N,h)$ with $S(h,N)$ to get the corresponding click
probability:
\begin{equation} 
t_S(R,N)=\int_{h_m}^{h_M}S(h,N)t(R,N,h)dh
\label{eq23}
\end{equation}
where $h_m$ and $h_M$ are the minimal and maximal fraction of pages in
a list, respectively.  We stress that if there is a maximal hit list
size $h_M<1$, each curve $t(R,N,h)$ of the overlap will have a flat
portion going from the first to the $1/h_{M}$-th page, so in the set of
pages ranked between $1$ and $1/h_M$ the click probability will be
flat, independently of the distribution function $S(h,N)$.

We obtained the hit list size distribution from a log of 200,000 actual
queries submitted to AltaVista in 2001 \linebreak (Fig.~\ref{fig3}B).  The data
can be reasonably well fitted by a power law with an exponential cutoff
due to the finite size of the AltaVista index.  The exponent of the
power law is $\delta \approx 1.1$.  In our Monte Carlo simulations we
neglected the exponential cutoff, and used the simple power law
\begin{equation} 
S(h,N) = B(N) h^{-\delta}
\label{eq23b}
\end{equation}
where the normalization constant $B(N)$ is just a function of $N$.  The
cutoff would affect only the part of the distribution $S(h,N)$
corresponding to the largest values of $h$, influencing a limited
portion of the curve $t_S(R,N)$ and the click probability of the very
top pages (cf.  the scaling relation of Eq.~\ref{eq22}).  As
there are no real queries that return hit lists containing all pages,%
\footnote{The policy of all search engines is to display at most $1000$
hits, and we took this into account in our simulations.  This does not
mean that $h \leq 1000/N$; the search engine scans all its database and
can report millions of hits, but it will finally display only the top
$1000$.} we have that $h_M<1$.  To estimate $h_{M}$ we divided the
largest observed number of Google hits in our collection of AltaVista
queries (approximately $6.6 \times 10^{8}$) by the total number of
pages reportedly indexed by Google (approximately $8 \times 10^{9}$ as
of this writing), yielding $h_M \approx 0.1$.  The top-ranked $1/h_M
\approx 10$ sites will have the same probability to be clicked.  We
then expect a flattening of the portion of $t_S(R,N)$ corresponding to
the pages with the highest PageRank/in-degree.  This flattening seems
consistent with the pattern observed in the real data
(Fig.~\ref{fig3}C).

\begin{figure}
    \centerline{\includegraphics[width=\columnwidth]{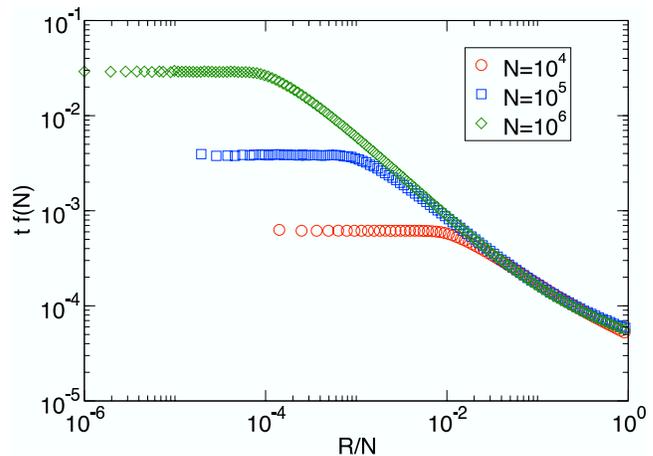}} 
    \caption{\label{figsuppl1}
    Scaling of $t_S(R,N)$ for $N=10^4, 10^5, 10^6$. The click probability $t$ is multiplied 
    for each curve by a number $f(N)$ that depends only on $N$. In the limit $N\rightarrow \infty$, $f(N)\rightarrow N$.}
\end{figure}

As to the full shape of the curve $t_S(R,N)$ for the Web, we performed
a simulation for a set of $N=10^6$ pages.  We used $h_m=1/N$, as there
are hit lists with a few or even a single result.  The size of our
sample is still very far from the total number of pages of the Web, so
in principle we could not match the curve derived from the simulation
with the pattern of the real data.  However, the theoretical curves
obey a simple scaling relation, as we can see in Fig.~\ref{figsuppl1}.
It is indeed possible to prove that $t_S(R,N)$ is a function of the
`normalized' rank $R/N$ (and of $N$) and not of the absolute rank $R$.
On a log-log scale, this means that by properly shifting curves
obtained for different $N$ values along the $x$ and $y$ axes it is
possible to make them overlap, exactly as we see in
Fig.~\ref{figsuppl1}.  This allows us to safely extrapolate to the
limit of much larger $N$, and to lay the curve derived by our
simulation on the empirical data (as we did in Fig.~\ref{fig3}C).
The argument is rather simple, and is based on the \emph{ansatz} of
Eq.~\ref{eq22a} for the function $t(R,N,h)$ and the power law form of
the distribution $S(h,N)$ (Eq.~\ref{eq23b}).  If we perform the
convolution of Eq.~\ref{eq23}, we have
\begin{equation} 
t_S(R,N)=\int_{1/N}^{h_M}S(h,N)h\,A(N)F(Rh)dh,
\label{eq23c}
\end{equation}
where we explicitly set $h_m=1/N$ and $F(Rh)$ is the universal function of Eq.~\ref{eq22}. 
By plugging the explicit expression of $S(h,N)$ from Eq.~\ref{eq23b}
into Eq.~\ref{eq23c} and performing the simple change of variable
$z=hN$ within the integral we obtain
\begin{equation} 
t_S(R,N)=\frac{A(N)B(N)}{N^{2-\delta}}\int_{1}^{h_M\,N}z^{1-\delta}\,F\left(\frac{R}{N}\,z\right)dz.
\label{eq24}
\end{equation}
The upper integration limit can be safely set to infinity because
$h_M\,N$ is very large.  The integral in Eq.~\ref{eq24} thus becomes a
function of the ratio $R/N$.  The additional explicit dependence on
$N$, expressed by the term outside the integral, consists in a simple
multiplicative factor $f(N)$ that does not affect the shape of the
curve (cf.  Fig.~\ref{figsuppl1}).

We finally remark that the expression $t_S(R,N)$ that we derived by
simulation represents the relation between the click probability and
the global rank of a page as determined by the value of its PageRank.
For a comparison with the empirical data of Fig.~\ref{fig3}C we need a
relation between click probability and in-degree.  We can relate rank
to in-degree by means of Eq.~\ref{eq10} between rank and PageRank and
by exploiting the proportionality between PageRank and in-degree
discussed earlier.

However both Eq.~\ref{eq10} and the proportionality between $p$ and $k$
are not rigorous, but only hold in the asymptotic regime of low
rank/large in-degree.  If it were feasible to simulate queries on a Web
graph with $O(10^{10})$ nodes, the theoretical curve in
Fig.~\ref{fig3}C would extend over the entire range of the x-axis.  In
this case the low-$k$ part of the curve would have to be adjusted to
account for the flattening observed in Fig.~\ref{figsuppl02}, which
displays the relation between PageRank and in-degree.  The leftmost
part of this curve is quite flat for over one order of magnitude,
giving a plausible explanation for the flat pattern of the low-$k$ data
in Fig.~\ref{fig3}C.

\balancecolumns 
\end{document}